\documentclass[12pt]{iopart}
\usepackage{graphicx}

\begin{document}

\title{Ontologies and tag-statistics}

\author{Gergely Tib{\'e}ly$^1$, P{\'e}ter Pollner$^2$, Tam\'as Vicsek$^{1,2}$ and Gergely Palla$^2$}
\address{$^1$ Dept. of Biological Physics, E{\"o}tv{\"o}s Univ.,
  1117 Budapest, P\'azm\'any P. stny. 1A}
\address{$^2$ Statistical and Biological Physics Research Group of HAS and}
\eads{\mailto{tibely@hal.elte.hu}, \mailto{pollner@hal.elte.hu}, \mailto{vicsek@hal.elte.hu}, \mailto{pallag@hal.elte.hu}}

\begin{abstract}
Due to the increasing popularity of collaborative tagging systems, the
 research on tagged networks, hypergraphs, ontologies, folksonomies and other
 related concepts is becoming an important interdisciplinary topic with great
 actuality and relevance for practical applications. In most collaborative 
tagging systems the tagging by the users is completely ``flat'', while in 
some cases they are allowed to define a shallow hierarchy for their own tags.
 However, usually no overall hierarchical organisation of the tags is given,
 and one of the interesting challenges of this area is to provide an algorithm
 generating the ontology of the tags from the available data. In contrast, there
 are also other type of tagged networks available for research, where the tags 
are already organised into a directed acyclic graph (DAG), encapsulating the 
``is a sub-category of'' type of hierarchy between each other. In this paper we
 study how this DAG affects the statistical distribution of tags on the 
nodes marked by the tags in various real networks. 
The motivation of this research is that 
understanding the tagging based on a known hierarchy can help in revealing
 the hidden hierarchy of tags in collaborative tagging systems. 
We analyse the relation between the tag-frequency and the position of 
the tag in the DAG in two large sub-networks of the English Wikipedia and 
a protein-protein interaction network. We also study the 
tag co-occurrence statistics by introducing a 2d tag-distance distribution
preserving both the difference in the levels and the absolute distance
in the DAG for the co-occurring pairs of tags.
Our most interesting finding is that the local relevance of tags in the DAG, 
(i.e., their rank or significance as characterised by, e.g., 
the length of the branches starting from them) is much more important than
their global distance from the root. 
Furthermore, we also introduce a simple tagging model based on
random walks on the DAG, capable of reproducing the 
main statistical features of tag co-occurrence.
This 
model has high potential for further practical applications, e.g., it can
provide a starting point for a benchmark system in ontology retrieval, or
it may help pinpointing unusual correlations in the co-occurrence of tags.
\end{abstract}

%PACS:
%02.70.Rr 	General statistical methods 
%89.20.-a 	Interdisciplinary applications of physics
%89.75.Hc 	Networks and genealogical trees 

%Keywords:
% networks, tags, folksonomy, ontology, hierarchy

%Suggested referees:
% Janos Kertesz kertesz@phy.bme.hu
% Santo Fortunato santo.fortunato@aalto.fi
% Alain Barrat Alain.Barrat@cpt.univ-mrs.fr
% Jari Saramaki jari.saramaki@tkk.fi

\section{Introduction}
The network approach has become an ubiquitous tool for analysing complex systems ranging from the interactions within cells through transportation systems, the Internet and other technological networks to economic networks, collaboration networks and the society \cite{Laci_revmod,Dorog_book}. Over the last
decade it has turned out that  networks corresponding to realistic systems
can be highly non-trivial, characterised by  a low average distance
combined with a high average clustering coefficient \cite{Watts-Strogatz},
anomalous degree distributions \cite{Faloutsos,Laci_science} and an intricate
modular structure \cite{GN-pnas,CPM_nature,Fortunato_report}. A recently emerging sub-field  of growing interest in this area is given by \emph{tagged networks, folksonomies} and \emph{hypergraphs}. In general, when studying the topology of the graph corresponding to a real system, the 
inclusion of \emph{node tags} (also called as attributes, annotations, 
properties, categories, features) leads
 to a richer structure, opening up the possibility for a more comprehensive
 analysis. These tags can correspond to 
 any information about the nodes and in most cases a single node
 can have several tags at the same time. The appearance of tags,
e.g., in biological networks is very common 
\cite{Mason_nets_in_bio,Zhu_nets_in_bio,Aittokallio_nets_in_bio,Finocchiaro_cancer,Jonsson_Bioinformatics,Jonsson_BMC}, 
where they usually refer to the biological function of the units represented
 by the nodes (proteins, genes, etc.). Node features are also fundamental ingredients in the so-called \emph{co-evolving} network models, where the evolution 
of the network topology affects the node properties and vice versa
\cite{Zimmermann_coevlov,Eguiluz_coevolv,Watts_science,Ehrhardt_coevolv,Newman_coevolv,Gil_coevolv,Vazquez_PRE,Vazquez_cond_mat,Kozma_coevolv,Benczik_coevolv,Castellano_coevolv}.
These models are aimed at describing
 the dynamics of social networks, in which people with similar opinion
 are assumed to form ties more easily, and the opinion of connected people
 becomes more similar in time. 

The entanglement between tags and the network structure is even more deep 
in \emph{collaborative tagging systems} or \emph{folksonomies} like CiteUlike, 
Delicious or Flickr \cite{Cattuto_PNAS,Lambiotte_ct,Cattuto_PNAS2}, where the 
network is actually arising in a tagging process. 
The basic scenario in these systems is
 that users can tag certain type of objects (photos, web-pages, books, etc.) 
with freely chosen words. Although the limits of the access to objects and tags introduced by others varies from system to system, the arising set of 
objects with associated free tags is usually referred to as a folksonomy. 
Since each tagging action is forming a new user-tag-object triple, 
the natural representation of these systems is given 
by tri-partite graphs, or in a more general framework by \emph{hypergraphs}
 \cite{Lambiotte_ct,Newman_PRE,Caldarelli_PRE}, where the hyperedges can 
connect more than two nodes together. In some cases the users are also offered 
the possibility to indicate social contacts (mark each other as a friend), 
opening up a new
dimension for the analysis of the interrelation between tagging and 
the social ties between  users \cite{Menzer_folk_soc,Scifanella_folk_soc}. 

Folksonomies provide an alternative
approach to organise knowledge compared to \emph{ontologies} \cite{Mika_folk_and_ont,Spyns_folk_and_ont,Voss_cond_mat}. An ontology usually corresponds to 
a set of narrower or broader \emph{categories}, (capturing the view 
and concepts of a certain domain, e.g., protein functions), building up
a hierarchy composed of ``is a sub-category of'' type relations. 
The natural representation of this hierarchy is given by 
a directed acyclic graph (DAG) between the categories.
 When tagging objects
with categories taken from an ontology, we have the benefit that
in principle all ancestors up to the root in the DAG can be inferred
from a single tag on the object.
 In contrast, the
tagging in a folksonomy is either completely ``flat'', or at most the users can
define a shallow hierarchy for their own tags. Nevertheless, a global 
hierarchical organisation of the tags is not given. 
One of the very interesting challenges related to folksonomies 
is to extract an ontology for the tags
appearing in the system. Several promising approaches have been
proposed, e.g., by aggregating the shallow hierarchies of the individual
users \cite{Lerman_constr,Lerman_constr_2}, using a probabilistic model 
\cite{Schmitz_constr}, analysing the node centralities in the co-occurrence
network between the tags \cite{Garcia-Molina}, or integrating 
information from as many sources as possible \cite{Van_Damme_constr}.
Since a reliable hierarchy between the tags can seriously improve searching,
 an effective ontology building algorithm has high 
potential for practical applications. 

Motivated by the ontology extraction problem described above, in this paper we
focus on the relation between the structure of the ontology and 
the distribution of the tags in systems where the DAG describing 
the hierarchical relations is predefined. 
The basic idea is that understanding how the ontology effects the tagging 
can help in improving the methods for reverse engineering the hidden 
DAG from the tag distribution in folksonomies. Along this line we examine 
the statistics of tag occurrence in two large sub-graphs of the English
Wikipedia and the protein interaction network of MIPS. We also analyse
 samples from Flickr, where the user defined shallow hierarchies are
 taken into account as individual DAGs. Furthermore, 
we introduce a simple model for reproducing the observed statistics
based on a random walk on the DAG of tags. The paper is organised
as follows: in Sect.\ref{sect:defs}. we define the most important
 quantities we aim to study, while the details of the investigated
 networks are given in Sect.\ref{sect:sys}. The obtained statistics are
presented in Sect.\ref{sect:app}, continued by the description of 
the random walk model in Sect.\ref{sect:model}, with some concluding
remarks closing the paper in Sect.\ref{sect:concl}.

\section{Definitions}
\label{sect:defs}

\subsection{DAG-levels}
In the tagged networks we study the tags are 
organised into a hierarchy which can be 
represented by a DAG, where the  directed links between two tags correspond
to an ``is a sub-category of'' type of relation. 
The tags close to the root in the DAG are usually related to
 general properties, and as we follow the links towards the leafs, the
 categories become more and more specific.
 In some cases we can find categories in the DAG with more than one 
in-neighbours, meaning that the given sub-category is part of several
 categories which are not in direct ancestor-descendant relation
with each other.

\begin{figure}[hbt]
\centerline{\includegraphics[width=0.6\textwidth]{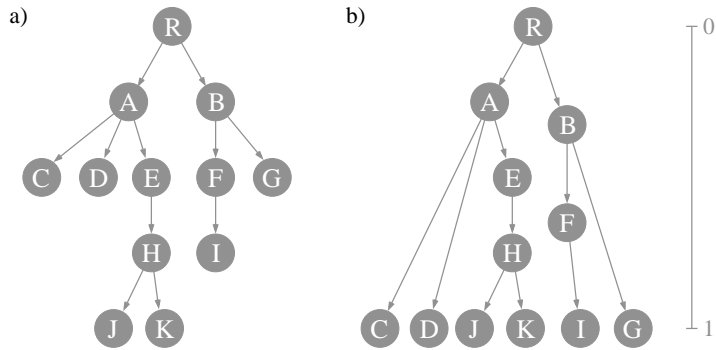}}
\caption{Illustration of the rescaling of the DAG. a) A small DAG of 
categories in which leaf nodes appear at various levels. The vertical position
of a tag (category) is determined by its distance from the root. b) After the 
rescaling the leafs are all at the bottom, and the vertical position
of each node is determined by the longest root-leaf path in which
it participates.}
\label{fig:haromszog_szeml}
\end{figure}

 Starting from the root we 
can define \emph{levels} in the DAG, with the root corresponding to 
level $l=0$, the first tags under the root providing level $l=1$, etc. 
For tags which can be reached via multiple paths from the root we assign the level corresponding to the 
longest path. (In some cases the level value of a tag is also referred to as
the rank of tag).
One of the simple statistical properties we are interested in
 is how does $l$ effect the frequency of the tags, 
or in other words, are the popular/rare tags close to the root 
in the DAG, or are they more likely to be close to the leafs? At this point 
we note that leafs can occur in principle at any level in the DAG, since the
different branches have usually different maximal depths in a real system.
In order to be able to judge the distance of a tag from the leafs as well,
 we introduce a rescaling of the level values illustrated in 
Fig.\ref{fig:haromszog_szeml}.:
The 
rescaled level value, $\tilde{l}$ at the root remains unchanged 
($\tilde{l}=0$), while for any leaf tags we require $\tilde{l}=1$. For
tags in-between the two extremes we assign an $\tilde{l}\in[0,1]$ based on the 
length of the longest root-leaf branch it takes part in, and 
$\tilde{l}$ is given by the depth of the tag divided by the maximal
depth of the branch. 

\subsection{Tag-frequency}
The \emph{frequency} of the tags in most real systems is heterogeneous, 
most popular tags occur rather often, whereas others are 
assigned only to a few objects. A natural choice for the definition
of the frequency $f_{\alpha}$ of a given tag $\alpha$ is simply the 
number of objects it is assigned to. The probability to find 
$\alpha$ attached to an object chosen uniformly at random is given by
\begin{equation}
p_{\alpha} =  f_{\alpha}/N,
\label{eq:freq}
\end{equation}
where $N$ denotes the total number of objects.
According to \cite{our_tagged}, the probability distribution of this 
quantity has a power-law like tail for both the Wikipedia and the
MIPS network. In the present work (among other questions) 
we shall be interested in
how is the tag-frequency effected by the level value of the tag.

We note that in systems where a DAG of the hierarchical relations between 
the tags is given, in principle we 
could infer all ancestors up to the root from an actually present
tag on an object. This enables an alternative definition of the 
tag-frequency \cite{our_tagged,our_tag_clust}, considering 
the aggregated number of occurrences for all descendants
of $\alpha$ and $\alpha$ itself. However, since one of the main motivation
of the present work is given by folksonomies (where the DAG is absent),
we shall concentrate on the frequency given by simply the number of occurrences.

\subsection{2d tag-distance distribution}
\label{def:map}

\begin{figure}[hbt]
\centerline{\includegraphics[width=0.6\textwidth]{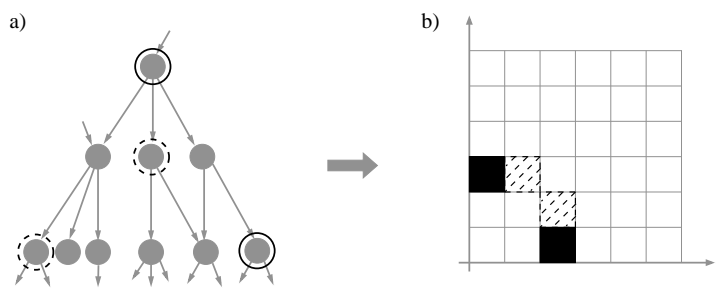}}
\caption{Illustration of the 2d tag-distance distribution for the 
 co-occurring tag pairs. a) A small part of a 
DAG with two pairs of tags chosen: the ones marked
with continuous circles are in direct ancestor-descendant relation, whereas
the tags marked with dashed circles form an ``uncle-nephew'' pair. b)
The corresponding cells of the tag-distance distribution are highlighted 
in solid black colour and with dashed-lines respectively.}
\label{fig:hoterkep_magy}
\end{figure}

Another question of interest is how does the DAG affect the co-occurrence
of different tags on the same object. The simplest idea for measuring
the relatedness of a pair of co-occurring tags based on the DAG 
would be given by their distance. 
However, for some pairs the connecting path 
in the DAG is composed of links all going in the same direction, whereas 
in other cases we might need both up- and downward pointing links to reach
from one tag to the other. In order to also include this aspect into the 
investigations, we define the \emph{2d tag-distance distribution} 
for the co-occurring pairs as illustrated in Fig.\ref{fig:hoterkep_magy}. 
The positive quarter plane 
is divided into unit cells, with the cell at the origin corresponding to
distance zero. A given pair of tags contributes to the distribution as follows: 
starting from one of them we 
first move upwards in the DAG until the lowest common ancestor is
reached. In parallel we move the same number of cells vertically up in the 
2d plane. Next, we move downwards in the DAG to reach the other
tag, and in parallel, we move the same number cells horizontally to
the right in the plane, and the number of ``events'' in the final cell 
is increased by one. The contribution from the path going back to the first tag 
from the second one
is taken into account following the same rules: going upwards in
the DAG corresponds to moving up in the 2d plane starting from
the origin, whereas going down in the 
DAG corresponds to moving horizontally to the right in the plane. 
The resulting distribution of the tag-distances is symmetric to the diagonal
by construction. The co-occurring pairs of tags which are in direct
ancestor-descendant relation contribute to the first column of cells and 
the bottom row, whereas e.g., the diagonal cells correspond to pairs
in which the two tags are equally deep in different branches from
their lowest common ancestor, 
(see Fig.\ref{fig:hoterkep_magy} for illustration).

\section{The studied systems}
\label{sect:sys}

We studied the statistical properties of co-occurring tags 
with predefined DAG in two sub-networks of the English Wikipedia and 
the protein-protein interaction network of MIPS. 
Furthermore, we also investigated the tag co-occurrence in
the presence of user defined shallow hierarchies in samples from Flickr.

The protein-protein interaction 
network of MIPS \cite{MIPS} consisted of $N=4546$ proteins, 
connected by $M=12319$ links, and the tags attached to the nodes corresponded 
 to $2067$ categories describing the biological processes the proteins
 take part in. The DAG between these categories was obtained from
 the Genome Ontology database \cite{GO}.

In the Wikipedia \cite{Eng_wiki} 
the pages are connected by hyperlinks (providing
a very interesting network on its own 
\cite{Eng_wiki,Zlatic_wikipedia,Capocci_wiki_PRE,Capocci_wikipedia}),
and at the bottom of each page, one can find
 a list of categories, which can be used as tags. We used the same
 data set as in \cite{our_tagged,our_tag_clust}, representing the
state of the system in 2008. Since each 
 Wiki-category is a page in the Wikipedia as well, these 
pages were removed from the network to keep a clear distinction between objects 
and tags. Similarly
 to the biological processes in the MIPS network, the Wiki-categories 
can have sub-categories and are usually part of a larger Wiki-category.
(Although the directed graph graph between the Wiki-categories 
contains a few loops, these can be removed quite easily to obtain
a strict DAG \cite{our_tagged}). 
Since the English-Wikipedia is quite a large network, we used 
smaller subsets obtained with a sampling method based on 
the tag-induced graphs \cite{our_tagged}: after choosing a rather
general category we keep the pages marked by this tag or any of its
descendants. The chosen sub-graphs were induced by the categories 
 ``Japan'', (consisting of $N=61581$ nodes, $M=949350$ links and
$4939$ sub-categories),  and ``United Kingdom'' 
(consisting of $N=318183$ nodes, $M=5432914$ links and $30383$ sub-categories).

One of the most popular collaborative tagging systems is 
given by Flickr, designed for tagging photos. Beside attaching
tags, the users can also group their photos into so-called sets, and
these sets can be also put into larger collections up to a 
limited range of levels. In contrast
to the photos, the collections and sets are given short descriptions 
rather than tags. Anyhow, from the sets and collections 
of a given user we can generate a shallow hierarchy of tags. A natural
choice is to link all tags appearing in a given set under one 
meta-tag corresponding to the set itself, 
then link this meta-tag under another meta-tag corresponding
to the collection the given set is part of, etc. Of course, the 
tag-distance distribution of the tags appearing on the photos of a given user
becomes trivial if we use the shallow hierarchy gained from these 
photos as the DAG: all co-occurring tags are siblings. However, the
picture becomes non-trivial when we calculate the tag-distance distribution of
the tags of a given user with help of the DAG gained from another
user. In fact, for a given sample of users from Flickr, we can
extract the shallow hierarchy of each user separately, then prepare
the tag-distance distribution of for all other users using this DAG, and finally
merge the results into one aggregated plot. 

\section{Applications}
\label{sect:app}

\subsection{The structure of the DAG}

\begin{figure}[hbt]
\centerline{\includegraphics[width=\textwidth]{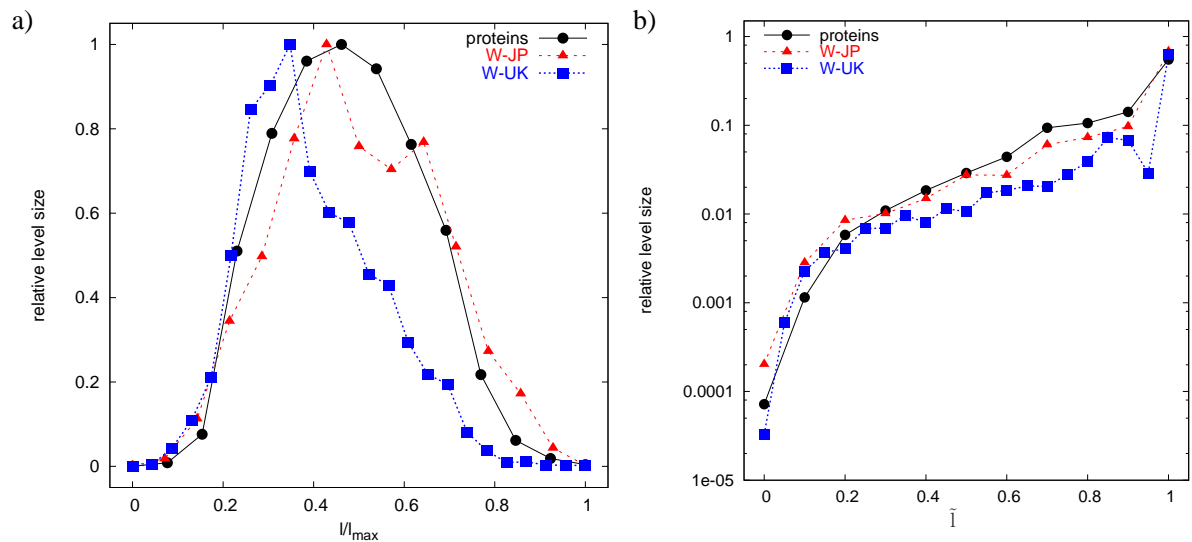}}
\caption{ a) The relative size of the levels in the DAG (scaled with the
largest level) as a function of $l/l_{\rm max}$, where $l_{\rm max}$ denotes the 
length of the longest branch in the DAG . b) The relative size of the levels
in the DAG as a function of the rescaled level depth $\tilde{l}$. Note that
the vertical axis is logarithmic.}
\label{fig:level_sizes}
\end{figure}

We start our analysis with an interesting effect related to
the structure of the DAGs describing the hierarchy of tags 
in the systems we investigate.
In Fig.\ref{fig:level_sizes}. we plot
the size of the levels (how many tags occur at a given level) as a
 function of the level depth.
For convenience, the vertical axis for each plot is rescaled by the 
size of the largest level. According to Fig.\ref{fig:level_sizes}a, 
the size of the levels is small when we are either very close to the root,
 or very far from it, whereas it becomes larger in between. 
(Since the maximal depth is different in each system, the horizontal axis in 
this case has been rescaled from $l$ to $l/l_{\rm max}$, 
where $l_{\rm max}$ denotes the length of the longest branch in the DAG). 
However, the shape and place of this maximum is unique for each system.
(An alternative illustration of this effect is given in the Appendix,
 where in the top-panel of Fig.\ref{fig:level_size_alt}. the differences
between the un-scaled DAGs are more apparent). 
In contrast,
 when we switch to the rescaled level depth $\tilde{l}$, the curves
become roughly uniform with a more or less monotonously increasing shape, 
as shown in 
Fig.\ref{fig:level_sizes}b.
Thus, the rescaling of the level values has an interesting side effect 
on the shape of the DAG, bringing it closer a ``triangular'' form, 
similar to the shape of a regular hierarchical graph.

\subsection{Tag-frequencies and level values}
\label{sect:tag-freq}

\begin{figure}[hbt]
\centerline{\includegraphics[width=\textwidth]{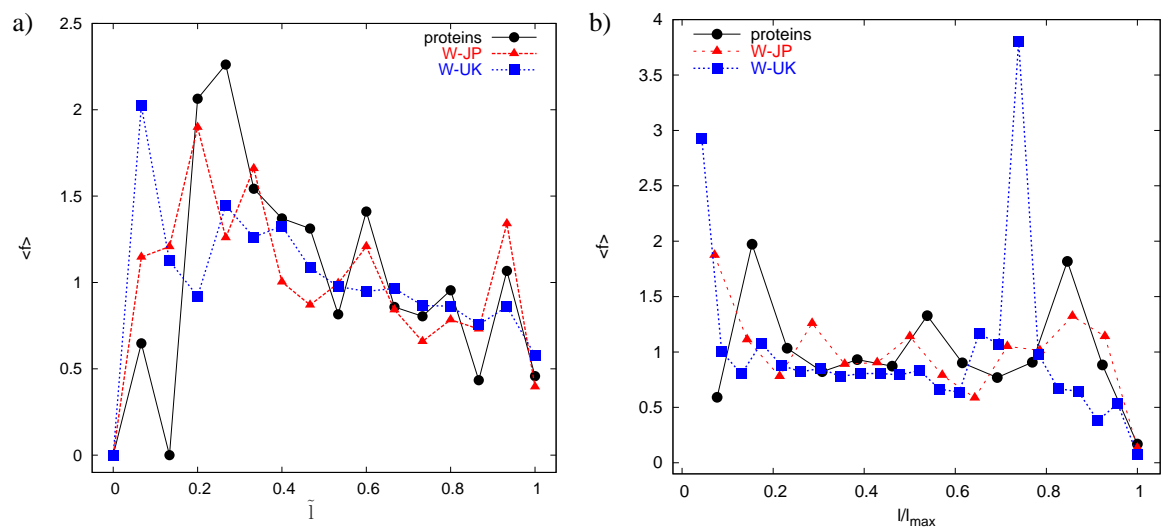}}
\caption{ a) The average frequency $\left< f\right>$ of the tags 
on a given level as a 
function of $l/l_{\rm max}$. b) The average frequency $\left< f\right>$ 
of the tags on 
a given level as a function of the rescaled level depth $\tilde{l}$.}
\label{fig:level_pop}
\end{figure}

As our main interest is focused on the interplay between the 
tag-hierarchy and the statistical properties of tag-occurrences,
in Fig.\ref{fig:level_pop}. we show the average tag-frequency as a function
of the level depth. When no rescaling is applied (apart from dividing
$l$ by the maximal level depth $l_{\rm max}$), the tag-frequency 
is almost completely independent of the level depth in a wide range of $l$ 
(Fig.\ref{fig:level_pop}a). In contrast, when switching to the 
rescaled $\tilde{l}$, a clear decreasing tendency can be observed, apart
from the very low $\tilde{l}$ region (corresponding to levels close to
the root). This non-trivial result indicates that 
the frequency of a tag is more sensitive to
the depth of the branches starting from it
compared to its
distance from the root.
A plausible explanation of this effect is the following:
 we already pointed out that leafs can
occur basically at any level in a large enough real world DAG. 
As we move upwards from the leafs,
presumably the importance (relevance, rank, significance, standing, etc.) 
of the tags is increasing at least in the first few steps. However,
since the leafs we started from were located at various levels, we arrive
to the conclusion that tags with higher relevance can also occur at
a wide range of levels in the DAG. Thus, the level value $l$ of a tag,
 measuring its global distance from the root
is not very informative in this
respect, and accordingly, it has no significant effect on the average
frequency of the tags. In contrast, by switching to the rescaled level value
$\tilde{l}$, we also take into account the depth of the local branches 
starting from the given tag, which seem to be more relevant for evaluating
the standing of a tag in the hierarchy, as the frequency of tags
is decreasing with $\tilde{l}$. (The more important tags have longer
sub-branches starting from them, thus, on average have lower $\tilde{l}$ 
values).

\subsection{Tag-distance and co-occurrence}
\label{sect:td_results}

\begin{figure}[hbt]
\centerline{\includegraphics[width=0.9\textwidth]{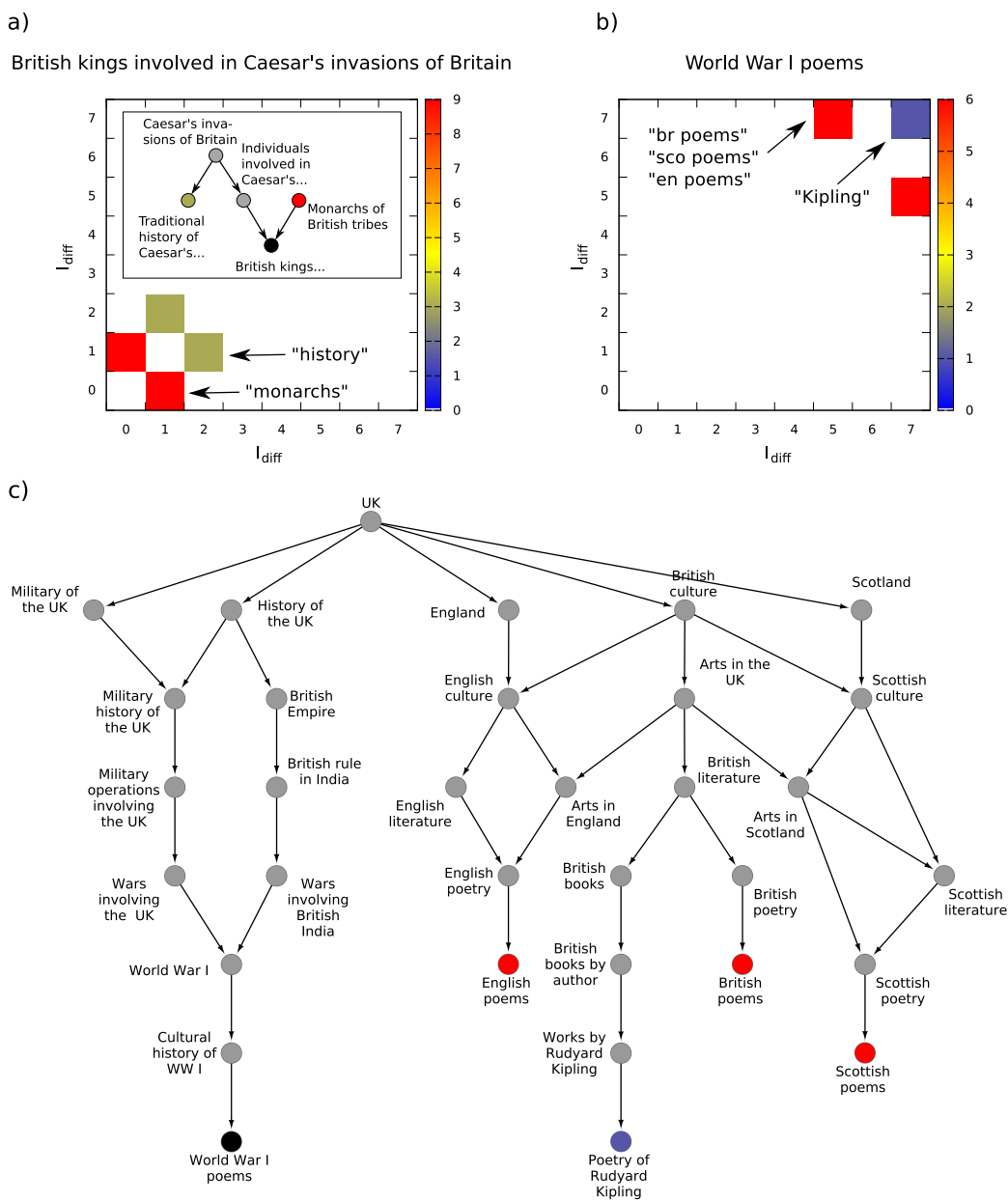}}
\caption{Illustration of the calculation of the tag-distance distribution
in case of the Wiki-UK network. 
a) The contribution from the tag 
``British kings involved in Caesar's invasion of Britain''. 
The colours indicate the number of occurrences, the inset shows the
corresponding sub-graph in the DAG.
b) The contribution from the tag ``World War I poems''. 
c) The shortest paths to the lowest common ancestors in
 the DAG for tag-pairs considered in panel (b).}
\label{fig:Caesar}
\end{figure}
Next we move on to the examination of the 2d tag-distance distributions
defined in Sect.\ref{def:map}. For illustration, in Fig.\ref{fig:Caesar}a
we show the contribution from the tag ``British kings involved in Caesar's 
invasion of Britain'' in case of the Wiki-UK network, 
where the number of occurrences together
with other tags at a given distance $l_{\rm diff}$ are indicated by the colour of the 
corresponding cell. The sub-graph between the tags and the lowest common
ancestors in the DAG is given in the inset. 
For comparison, in Fig.\ref{fig:Caesar}b we show 
the contribution from 
``World War I poems''  in a similar fashion. 
Here the routes through the lowest common ancestors between
the co-occurring tag pairs are much longer,
thus, they are displayed separately in Fig.\ref{fig:Caesar}c.
The tags co-occurring with ``British kings ...'' 
are close in the DAG, and accordingly,
their contribution in the 2d distribution is close to the origin. In contrast,
the distances to the tags co-occurring with ``World War I poems'' 
are large, thus,
their contribution falls in cells far from the origin.

\begin{figure}[hbt]
\centerline{\includegraphics[width=\textwidth]{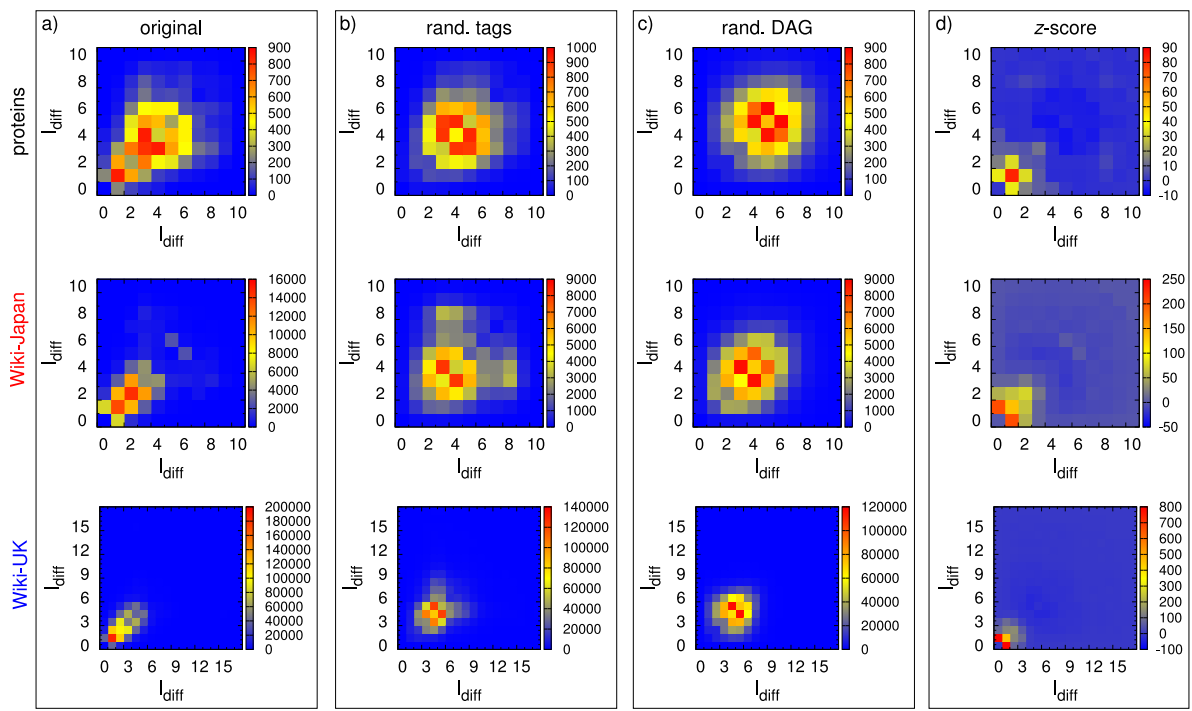}}
\caption{a) The 2d tag-distance distribution for co-occurring tag pairs in 
the studied systems (colour coded).
b) The average tag-distance distribution when the tags are randomised keeping the tag-frequencies and the number of tags on the objects fixed. 
c) The average tag-distance distribution for random DAGs. 
d) The $z$-score corresponding to the difference between the original
data (panel a) compared to the random tag assignment (panel b) in the units of
the standard deviation of the random tag assignment.}
\label{fig:heat_map_result}
\end{figure}
In Fig.\ref{fig:heat_map_result}. we plot the complete 2d tag-distance 
distributions for
the networks we investigated. According to the Fig.\ref{fig:heat_map_result}a, 
the maximum of the plots is a few steps away from the origin, 
which might seem a bit surprising at first sight. In order to reveal the
background of this effect we also measured the average 
tag-distance distribution for a random tag assignment analogous to
the configuration model in the networks literature. Here the DAG is 
taken from the system under study, and we
consider the ensemble of all possible associations of tags to the
objects consistent with the observed number of occurrences for the tags
and the observed number of tags on the individual objects.  
To simulate draws from this ensemble one can apply a randomisation
procedure, in which a pair of tags is swapped between two 
randomly chosen objects in each step. This way both the 
number of tags on the objects and the tag-frequencies are preserved. 
 The average tag-distance distributions for this random tag assignment 
are shown
 in Fig.\ref{fig:heat_map_result}b, with the maximums even further away from the
origin compared to the original data. An alternative possibility 
for randomisation
is to replace the DAG of the original system by a random DAG of the
the same size. For this we used the random DAG model introduced in
 \cite{Newman_random_DAG} with fixed number of nodes and links. The
 results for random DAGs and the original distribution of the tags 
on the objects are displayed in Fig.\ref{fig:heat_map_result}c, 
showing a picture somewhat similar to Fig.\ref{fig:heat_map_result}b.
Finally, to highlight the part of the 
original tag-distance distribution that cannot be accounted for 
random effects, in Fig.\ref{fig:heat_map_result}d we show 
the $z$-score of the individual cells, defined as 
the  difference between the original distribution (Fig.\ref{fig:heat_map_result}a) and the average for the random tag assignment (Fig.\ref{fig:heat_map_result}b) scaled by the standard deviation of the random tag assignment. 
The maximums in this plots have clearly moved close to the origin, 
showing that the co-occurrence of
tags only a few steps away in the DAG is far more probable than at random in
the systems we investigated.

\begin{figure}
\centerline{\includegraphics[width=\textwidth]{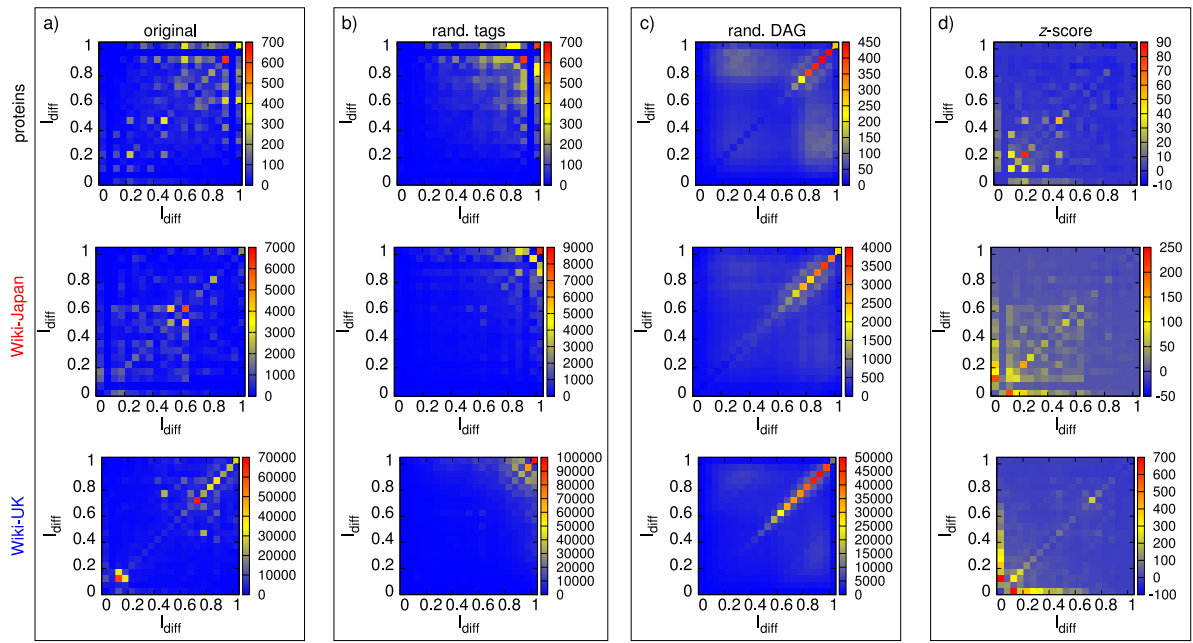}}
\caption{ The tag-distance distributions shown 
in Fig.\ref{fig:heat_map_result_resc}. when the distances are measured 
according to the  rescaled level value $\tilde{l}$. 
Since $\tilde{l}$ can take up real values (not only integers as $l$), 
we introduced bins of size 0.05. Similarly to 
Fig.\ref{fig:heat_map_result_resc}., beside the original data (panel a) we
also show the results for random tag assignment (panel b), random DAGs 
(panel c), and the $z$-scores (panel d).}
\label{fig:heat_map_result_resc}
\end{figure}
The three $z$-score plots in Fig.\ref{fig:heat_map_result}d also reveal an interesting difference
between the systems: in case of the protein interaction network the 
maximum is in the diagonal, while for the two Wiki-networks it is 
in the first row (or first column). This means that for the protein
 interaction network the most enhanced co-occurring tags are like 
``brothers'', i.e., they are at the same depth from their lowest common
ancestor on different branches. In contrast, the maximum places for
the Wiki-networks correspond to tag-pairs in direct ancestor-descendant
relation with each other. This effect is somewhat even more apparent when
we replot the tag-distance distributions using the rescaled level value 
$\tilde{l}$ for measuring the distance between the tags, as shown in
Fig.\ref{fig:heat_map_result_resc}.
(Since $\tilde{l}$ can take up real values in $\left[ 0,1\right]$, 
we introduced bins of size 
0.05 when preparing the tag-distance distributions). 
Similarly to Fig.\ref{fig:heat_map_result}., the maximums are far from
the origin for both the original data sets (panel a) 
and their random counterparts (panels b-c).
 However, in case of the $z$-score (panel d),
the maximum is shifted rather close to the origin along the diagonal
for the protein interaction network, while it is concentrated in
the first row (or column) for the two Wiki-networks. 
A further nice feature of using the rescaled levels is that the tag-distance
distribution of the random tag assignment has high values around (1,1), 
in contrast to the traditional level-based distribution which has high values in a nontrivial, case-specific region (see Fig.\ref{fig:heat_map_result}b).

\subsection{Results for Flickr}
\begin{figure}
\centerline{\includegraphics[width=\textwidth]{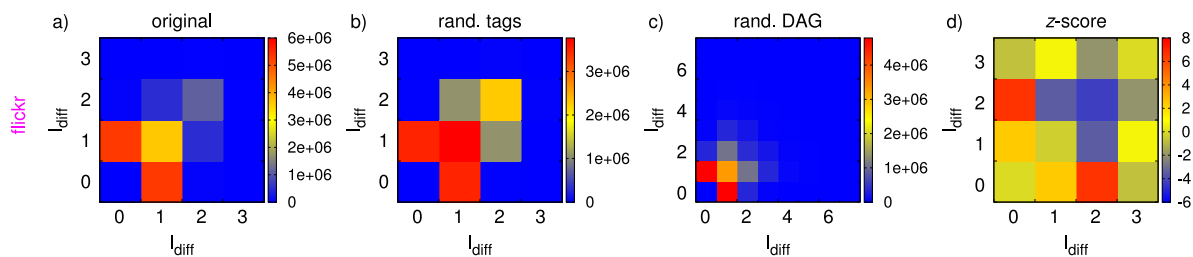}}
\caption{a) The tag-distance distribution obtained for a sample
from Flickr. We measured the distances between the co-occurring tags on a given 
photo belonging to a given user using the shallow hierarchies of the 
other users in the sample, and aggregated the results for all photos 
and all DAGs. b) The average tag-distance distribution for random
tag assignment. c) The average tag-distance distribution for random DAGs d) 
The $z$-score obtained by comparing the results shown in panel a) and the
null-model displayed in panel b).}
\label{fig:flickr}
\end{figure}

We also prepared the 2d tag-distance distribution using the
user defined shallow hierarchies for a sample from Flickr. 
Since the maximal level depth in this case is only $l=3$, 
the distinction between close by and far away tags becomes
a bit artificial, (e.g., for direct
descendants the largest possible distance is 3).
According to the results shown in 
Fig.\ref{fig:flickr}., the maximum of the 2d tag-distance distribution is 
close to the origin for both the original data (Fig.\ref{fig:flickr}a) 
and its randomised counterparts (Figs.\ref{fig:flickr}b-c). 
From the $z$-score (Fig.\ref{fig:flickr}d) we can see that
the cells having the most significant enhancement in the number of
tag-pairs compared to the random tag assignment correspond to
direct descendants within distance 1 and 2. 
Although this behaviour is consistent with the results shown previously for the 
three tagged networks,
the enhancement in the number of close by tag-pairs is far less striking. 
An interesting question, (which is out of the scope of the present work), 
related to the above is the following:
How would the tag-distance distribution 
behave in the Flickr data set if the set of user-defined shallow hierarchies 
are replaced by a unique overall DAG obtained from an ontology 
extraction algorithm?

\subsection{Local standing vs. global rank in the tag-distance distribution}
\label{sect:upper_lower_halfs}

In Sect.\ref{sect:tag-freq}. we have seen that
the length of the local branches starting from a given tag in the 
DAG have much larger effect on its frequency compared to its global 
distance from the root.
An interesting question related to this is whether we can observe any similar
effect in the behaviour of the 2d tag-distance distribution as well. 
Since this distribution depends on the relative distance between the 
co-occurring tags, from its original form we cannot deduce any information
about the absolute level value of the occurring tags. However, 
applying a randomisation restricted to a given part (e.g., a given set
of levels) in the DAG and tracking the induced changes in the 
2d tag-distance distribution can help resolving the influence of 
the given region in the DAG on the tag co-occurrence: If the chosen part 
is crucial, then the behaviour of 2d tag-distance distribution obtained
after the randomisation should be very different from the original. However,
 if the chosen part of the DAG has only little influence on the 
tag co-occurrence, then the restricted randomisation of the given part
should not make a significant difference.

\begin{figure}
\centerline{\includegraphics[width=\textwidth]{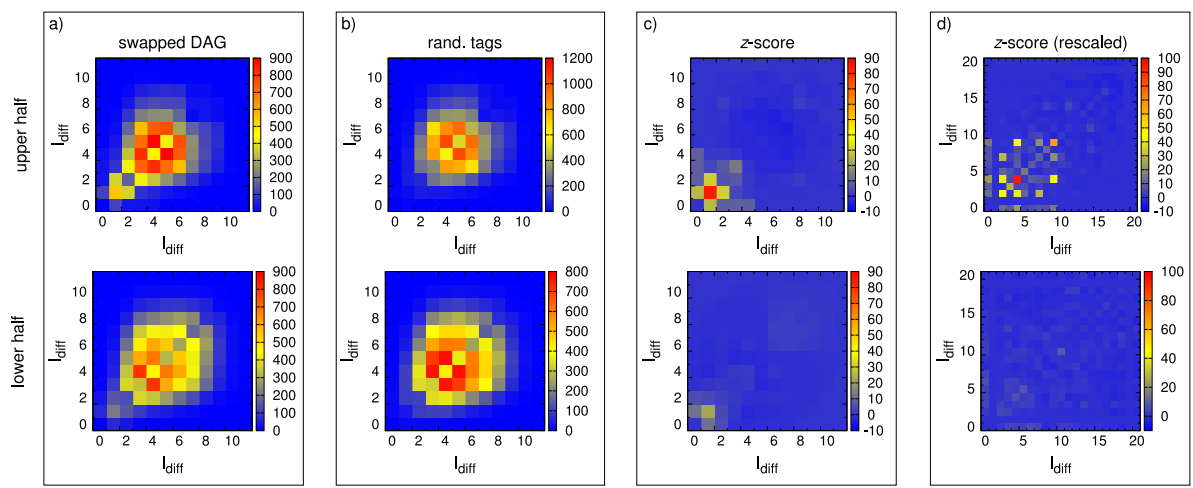}}
\caption{
Comparison between the tag-distance 
distributions obtained after randomisation of the ``upper half'' of the DAG
(top row) and the ``lower half'' of the DAG (bottom row) 
in case of the protein interaction network.
a) The obtained 2d tag-distance distributions. 
b) The average 2d tag-distance distributions for random tag assignment  
 (using the partially randomised DAGs). c) The $z$-score corresponding
to the difference between panel a and panel b in the units of the
standard deviation of panel b. d) The $z$-score when the distance
between the tags is measured according to the rescaled level 
value $\tilde{l}$.}
\label{fig:half_dag_rand}
\end{figure}

Along this line we divided the DAG of the systems we investigated into
an ``upper half'' (corresponding to levels close to the root), and
a ``lower half'' (composed of bottom levels far from the root).
Although the number of tags in the two parts were the same, the 
induced changes in the 2d tag-distance distribution due to the randomisation
of a single part alone were strikingly different, 
in agreement with the previously
observed enhancement of the importance of the ``local position''
compared to the``global position''  in the DAG from the point of view of 
tag frequencies. In Fig.\ref{fig:half_dag_rand}. we show the results
for randomising the ``upper half'' (top row) 
and ``lower half'' (bottom row) of the DAG
separately in the protein interaction network. 
(During the randomisation process at
each step a pair of tags from the restricted set was randomly swapped 
in the DAG). 
In Fig.\ref{fig:half_dag_rand}a we show the 
2d tag-distance distribution obtained after the restricted randomisation, 
for comparison Fig.\ref{fig:half_dag_rand}b displays the 
results for random tag assignment (using the partly randomised DAG). 
The corresponding $z$-scores are given in Fig.\ref{fig:half_dag_rand}c-d,
for both for the original- and the rescaled level values. 
 In case of randomising the ``upper half''  the $z$-scores are quite
similar to the $z$-scores shown in 
Figs.\ref{fig:heat_map_result}-\ref{fig:heat_map_result_resc}, (although
some small details look slightly different). In contrast, the $z$-scores
for randomising the ``lower half'' show drastic deviations from
the original $z$-scores: 
we can observe only a week reminiscent of the maximum close to the origin, 
and the landscape becomes almost completely flat. 
This enhanced sensitivity of the 2d tag-distance distribution
to the changes in the ``lower half'' of the DAG compared to changes
in the ``upper half'' is in agreement with the enhanced sensitivity
of the tag-frequencies to the local position of the tags in the 
hierarchy compared to the global distance from the root: 
When randomising the ``upper half'', the local position for at least the 
tags in the ``lower half'' is preserved, whereas the global routes to any
tag are messed up. In contrast, when randomising the ``lower half'',
while preserving the global structure, we mess up the local position for
the majority of the tags, (as ``upper half'' tags are also likely to have
branches reaching into the ``lower half''). 
We observed  similar behaviour in case of 
randomising partly the DAG of either the Wiki-Japan or the Wiki-UK
 network as well.

\section{Random walk model}
\label{sect:model}

According to the results  of Sect.\ref{sect:app}. the DAG between the tags
has indeed an effect on the co-occurrence of tags.
In this section we demonstrate that a rather simple model 
can reproduce the main statistical features observed for the real systems. 
Since the co-occurring tags were more close to each other in the DAG
than at random, the model has to provide a mechanism for choosing pairs
from the DAG with an enhanced probability for close by tags. A natural 
idea is to pick the first tag at random, then start a random
walk on the DAG from the chosen tag, and after a few steps pick the 
target reached. (In some respect this approach is a sort of ``dual-model'' of
the random walk model introduced in \cite{Arenas_words} on the network of 
word associations, which was used for inferring 
similarity relations between words).

If we take the DAG as predefined (e.g., the DAG of the system we would
like to model), then the two ``parameters'' of the model are given by 
the frequency distribution
of the tags and the length distribution of the random walks. 
For the 
tag-frequencies the first natural idea is to use the frequency
distribution measured in the real data. However, we also worked with
uniform tag-frequencies set to the average value measured in the real data.
In case of the random walk length distribution we tried
 out the gamma-distribution, the uniform distribution,
the lognormal distribution, and the Poisson-distribution. For all choices
the average length of the walks was set to a value ranging between 3 and 10. 
According to the results, the 
tag-distance distribution is very robust against the changes in any parameters.

\begin{figure}[hbt]
\begin{center}
\includegraphics[width=\textwidth]{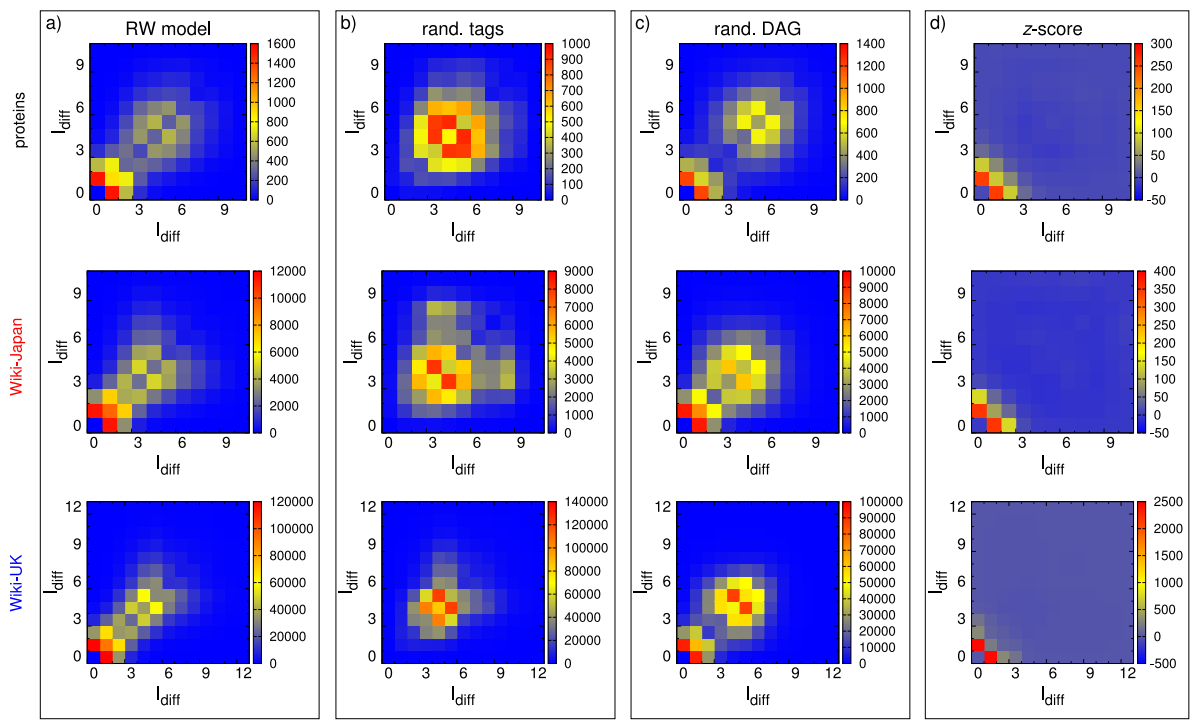}
\end{center}
\caption{ The tag-distance distributions for co-occurring pairs of tags in 
the random walk model (colour coded), where the DAG was taken from the 
protein interaction network (1$^{\rm st}$ row), the Wiki-Japan network 
(2$^{\rm nd}$ row) and the Wiki-UK network (3$^{\rm d}$ row). Similarly
to Fig.\ref{fig:heat_map_result}., beside the actually measured
values (panel a), for comparison 
the results for un-correlated tag assignment (panel b), 
the results for random DAGs (panel c)
and the z-scores (panel d) are also shown.}
\label{fig:heat_map_rand_walk}
\end{figure}

In Fig.\ref{fig:heat_map_rand_walk}a we show typical 
tag-distance distribution results
for the random walk model. The DAGs used in these
simulations (indicated on the left of each row) were 
taken from the tagged networks we studied in Sect.\ref{sect:td_results},
 and the frequency of the tags as well as the number of tags on 
the objects were set to the average values measured in the corresponding
 real system. The length distribution for the random walks was 
a uniform distribution in the [3-10] interval. 
(In the Appendix we show very similar results for different random walk
length distributions).
 Similarly to the case 
of the real systems we studied, 
in Fig.\ref{fig:heat_map_rand_walk}b we also show the results for 
a suitably chosen null-model which in this case
 corresponds to choosing the tag-pairs at random, irrespectively of the DAG.
In Fig.\ref{fig:heat_map_rand_walk}c we show the results for random walks
on random DAGs generated using the model introduced in \cite{Newman_random_DAG}.
For highlighting the part which is only present due to the correlations
induced by the random walk, in Fig.\ref{fig:heat_map_rand_walk}d we
also display the $z$-scores (corresponding to the difference between
Fig.\ref{fig:heat_map_rand_walk}a and Fig.\ref{fig:heat_map_rand_walk}b in
the units of the standard deviation of Fig.\ref{fig:heat_map_rand_walk}b). 
Similarly to the behaviour observed
in the real systems,  the maximum in the $z$-score is shifted
close to the origin in all cases.

In summary, according to the simulations, our
random walk model qualitatively reproduces the main properties of the 
tag-distance distribution for the co-occurring tags observed in real systems. 
Although the model is rather simple, it has high potential for further 
applications. On the one hand, with the further development of ontology
extraction algorithms sooner or later the need for a controllable benchmark
system will arise. The rough outline of this benchmark is that
stochastic collections of tagged ``objects'' (sets of co-occurring tags) 
are generated based on given input DAG, and using these collections 
as input we can test how well does a given
ontology extraction algorithm recover the DAG. Our random walk model can
provide a starting point for generating random collections of co-occurring
tags with some sort of ``memory'' of the underlying DAG of hierarchy 
between the tags. 

On the other hand, the tag-distance distribution 
generated by the random walk model can also help 
pointing out non-trivial effects in the tag-distance distribution of
the original data where the DAG was taken from. 
For example,  when using the DAG taken from the Wiki networks, 
the behaviour of the tag-distance distribution in the random walk model 
is very similar compared to the original one. However, when
repeating this experiment with the protein interaction network,
the two tag-distance distributions are only roughly similar: 
although the maximum in
the $z$-scores is shifted close to the origin in both cases,
for the original data it remains in a diagonal
position (see Figs.\ref{fig:heat_map_result}-\ref{fig:heat_map_result_resc}),
 while it becomes off-diagonal for the random walk model
(see Fig.\ref{fig:heat_map_rand_walk}). Thus, the co-occurrence patterns
of the protein interaction network have features which cannot be explained
by a simple random walk on the DAG.

\section{Conclusions}
\label{sect:concl}
Motivated by the ontology extraction problem in collaborative tagging systems
and folksonomies, we studied the statistical properties of tag occurrence
in tagged networks where the DAG of hierarchy between the tags is predefined. 
In order to be able to give support for the further development of 
ontology extraction algorithms, this research was focused on the 
interaction between the DAG and the tag-statistics. 
Our most interesting result is that the local standing (rank, significance,
etc.) of the tags in the DAG 
has a much more relevant effect on the tag-statistics
compared to the global distance from the root. 
This is supported on the one hand by the change in the behaviour 
of the tag-frequency
as a function of the level value when switching to the rescaled 
levels $\tilde{l}$, on the other hand by the different sensitivity
of the 2d tag-distance distribution for co-occurring tags to randomising
the ``upper half'' or the ``lower half'' of the DAG.

According to our studies on a protein interaction
network and two
sub-networks from the English Wikipedia, the average 
frequency of the tags is more or less independent of the
level value  
(distance from the root) in the hierarchy of the tags.
In contrast, if we switch to a rescaled level value
 $\tilde{l}$ taking into account also the length of the sub-branches
starting from the given tag in the DAG, we see a decreasing tendency in
the tag-frequency with growing $\tilde{l}$ in a wide range of $\tilde{l}$. 
A plausible explanation for this interesting effect is that the
distance from the root is not a good indicator of the importance
 (significance, rank, etc.), e.g., in the DAGs we studied leafs 
(corresponding probably to the most specific tags) occurred at 
a wide range of levels. However, the lengths of the branches starting from
a given tag provide an alternative candidate for evaluating its importance,
 and in contrast to the distance from the root, 
this  measure is of local nature.
The above result suggests that taking into account this local information as 
well when evaluating the rank of a tag yields a quantity which is much more
entangled with the tag-frequency compared to the traditional level value.

We studied the statistical properties of co-occurring tag pairs 
by introducing a 2d tag-distance distribution for the relative
positions in the DAG. We compared this distribution for the three investigated
systems with the distribution obtained for a random tag assignment analogous to
the configuration model in the complex network literature. According 
to the $z$-scores, close by pairs of tags co-occur in these systems far more
 often then expected at random. Furthermore, these 2d plots also reveal 
an interesting difference between the protein interaction network and
the Wiki-networks: in the first system the co-occurring tag pairs are much 
less likely to be direct descendants of each other compared to the
other two networks, instead they are often to like ``cousins'', ``brothers'' 
or ``nephews''. 
We also analysed the 2d tag-distance distribution obtained
for a sample from Flickr using the shallow hierarchies defined by the users.
The results were consistent with the behaviour seen for the tagged networks
with predefined DAG, however the increase in the number of 
the close by tag pairs compared to the random null model was far less 
striking.

In order to examine the difference between importance of the local- and global
position of the tags in the hierarchy from a further perspective, we
applied restrictive randomisation to the DAG by dividing it into
an ``upper-half'' and a ``lower-part'' of equal size. The induced changes
in the 2d tag-distance distribution showed significant difference:
the effect of randomising the ``upper-half'' is marginal, whereas the 
structure of the $z$-score undergoes a drastic transformation when
randomising the ``lower-half''. Since randomising the ``upper-half''   
 modifies mainly the global structure, while randomising 
the ``lower-half'' reshuffles mainly the local structure, this effect 
is in complete agreement with
the previously observed imbalance between the importance of the 
local- and global standing of tags (in favour of the local one) 
from the point of view of tag-frequencies.

Finally, we introduced a simple model based on random walks on the DAG 
for describing the enhancement of close by tag-pairs in the tag-distance
distribution. According to our simulations, this approach can reproduce 
the shift of the maximum towards the origin in the $z$-score in a robust way.
Although simple in nature, this model has relevant potential for further
applications, e.g., it can provide a starting point in constructing 
benchmark systems for ontology extraction algorithms, and can also help
in pinpointing non-trivial effects in the tag-distance distribution of
 real systems.

\section*{Appendix}
\label{sec:app}

\subsection*{A1: The structure of the DAGs}

\begin{figure}[hbt]
\begin{center}
\includegraphics*[width=0.6\textwidth]{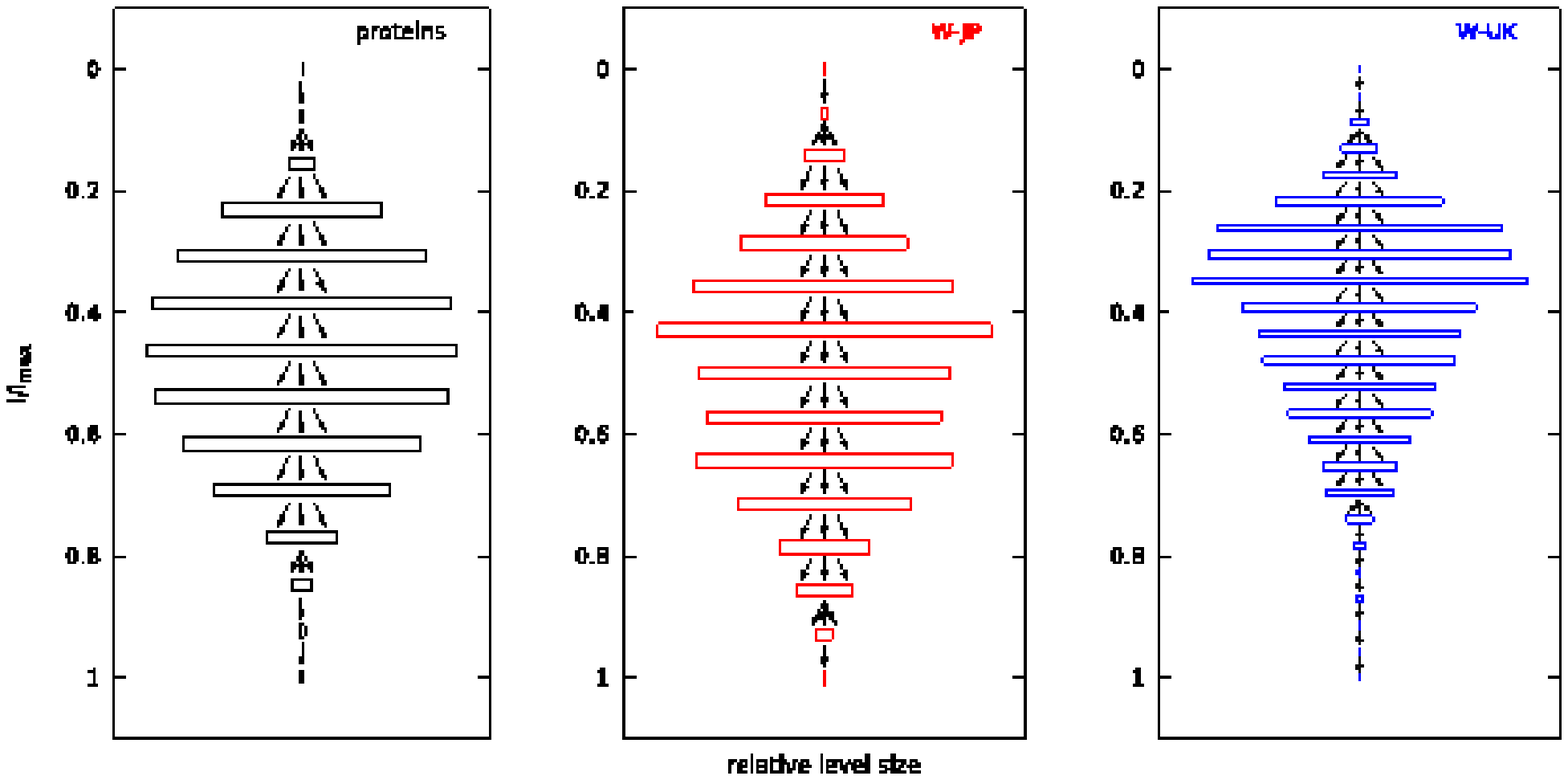}
\includegraphics*[width=0.6\textwidth]{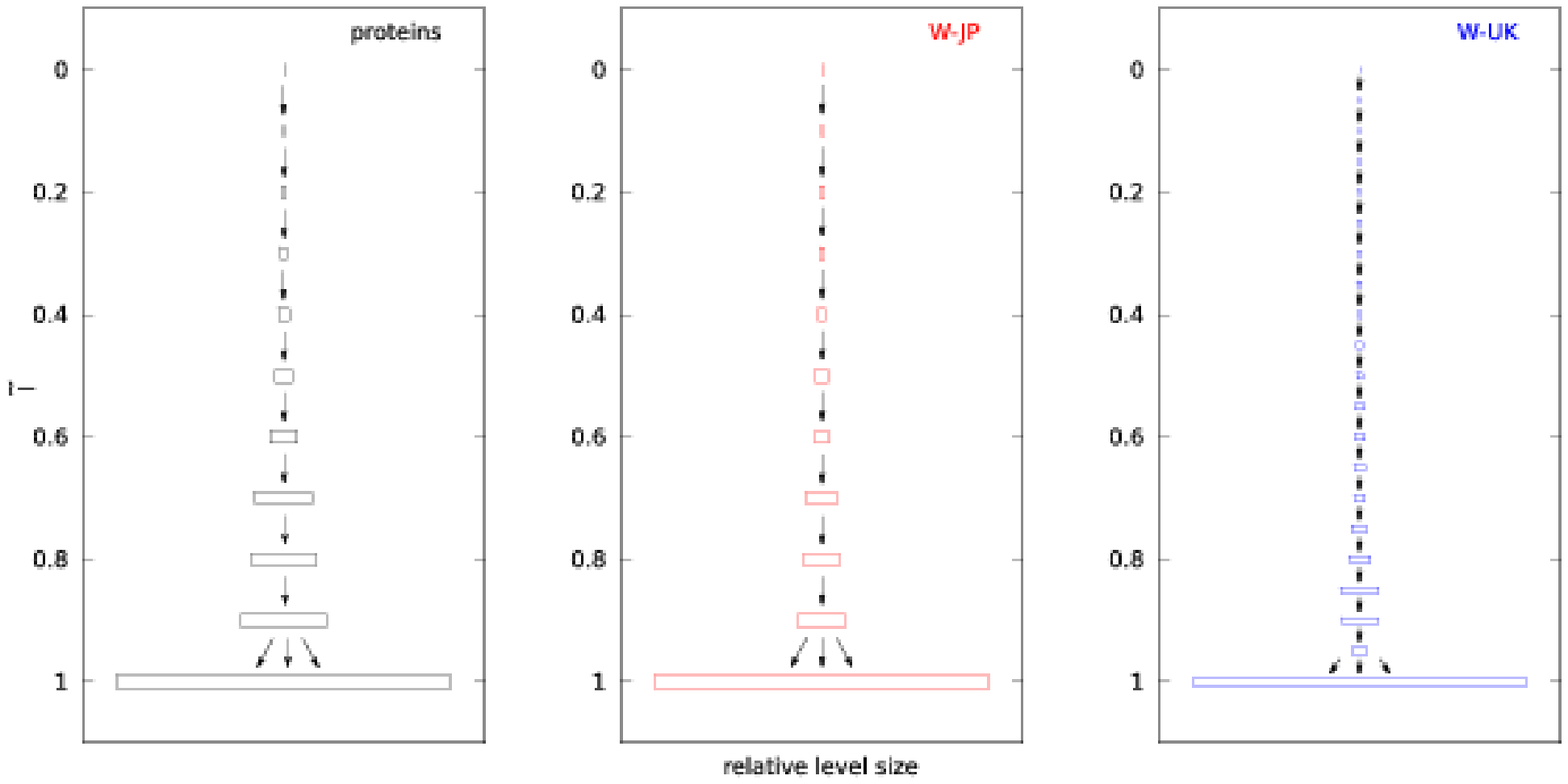}
\end{center}
\caption{a) Schematic illustration of the DAGs in the tagged networks we
 investigate. The width of the bars corresponds to the number of tags
at the given level. b) After switching to the rescaled level value 
$\tilde{l}$ the shape of the DAGs becomes rather uniform. }
\label{fig:level_size_alt}
\end{figure}
The DAG capturing the hierarchical relations between the tags plays
 a crucial role in our analysis, and in the systems we investigate the 
structure of the DAG is not trivial, i.e., its shape is far from e.g., the 
shape of a regular hierarchical graph in which the level sizes are
increasing as a power-law with the level depth. 
In Fig.\ref{fig:level_size_alt}a we show a schematic illustration of the
level sizes for the networks under study, where the width of the bars 
indicates the number of
tags on a given level, while the vertical position of the bar corresponds
to $l$. This representation shows significant differences between the
three DAGs. However, when switching to the rescaled level value $\tilde{l}$,
according to Fig.\ref{fig:level_size_alt}b, the shape of the DAGs become more or
less uniformly ``triangular''. (Since $\tilde{l}$ can take up real values
instead of integers, we used binning similarly to the case of 
Fig.\ref{fig:heat_map_result_resc}. in the main text).

\subsection*{A2: Robustness of the random walk model}

\begin{figure}[hbt]
\begin{center}
\includegraphics[width=\textwidth]{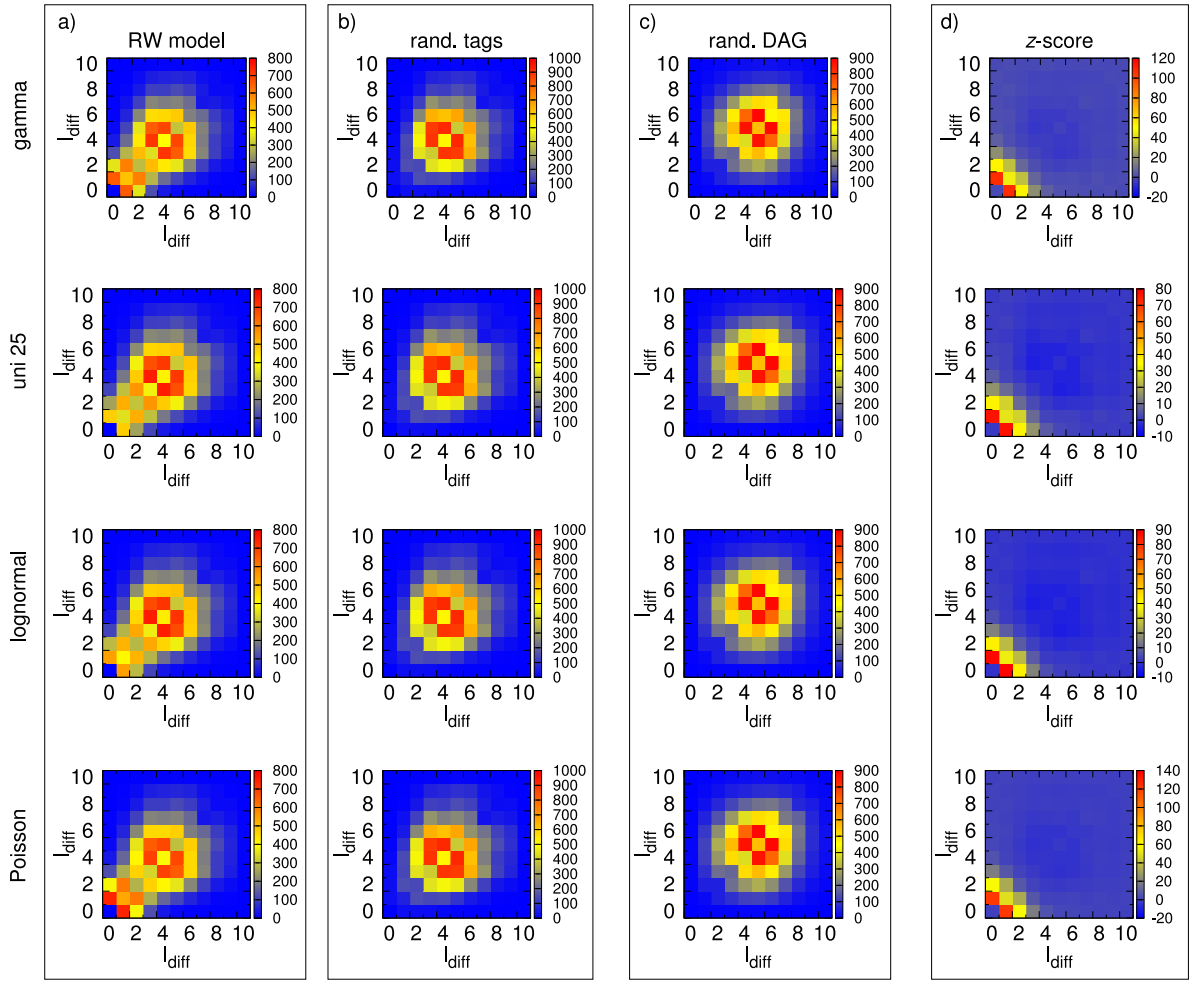}
\end{center}
\caption{The 2d tag-distance distribution in the random walk model when 
changing the walk length distribution to gamma-distribution (1$^{\rm st}$ row),
uniform distribution (2$^{\rm nd}$ row), lognormal-distribution
 (3$^{\rm d}$ row), and Poisson-distribution (4$^{\rm th}$ row) 
for the DAGs taken from the protein interaction network. 
Similarly to Fig.\ref{fig:heat_map_rand_walk}.,
beside the actually measured
values (panel a), for comparison 
the results for un-correlated tag assignment (panel b), 
the results for random DAGs (panel c)
and the $z$-scores (panel d) are also shown.}
\label{fig:heat_map_rand_walk_app}
\end{figure}

As mentioned in the main text, the random walk model turned out to be
quite robust against changes in the details like the frequency distribution
of the tags, the distribution of the number of tags on the objects, or
the length distribution of the random walk on the DAG. For illustration, 
in Fig.\ref{fig:heat_map_rand_walk_app}. we show results for replacing the
uniform distribution of the random walk lengths in 
Fig.\ref{fig:heat_map_rand_walk}. in the main text by gamma-distribution (1$^{\rm st}$ row),  uniform distribution with different ranges (2$^{\rm nd}$ row), 
lognormal-distribution (3$^{\rm d}$row), and Poisson-distribution 
(4$^{\rm th}$ row).
Apparently, the qualitative behaviour of the 2d tag-distance distribution is 
the same as before: the maximum is shifted close to the origin 
in the $z$-score. 

\section*{Acknowledgment}
This work was supported by the European Union and co-financed by the
European Social Fund (grant agreement no. TAMOP 4.2.1/B-09/1/KMR-2010-0003).

\section*{References}

\bibliographystyle{unsrt}
\bibliography{fgm_art}

\end{document}